\documentclass[11pt]{article}

\usepackage{comment}

\usepackage{amssymb}
\usepackage{amsmath}
\usepackage{amsthm}

\usepackage{fullpage}  
\usepackage{color}  
\usepackage[linesnumbered,noline,boxed,ruled,algosection]{algorithm2e}
\usepackage{hyperref}

\SetKwInput{Input}{input}
\SetKwInput{Output}{output}
\SetAlFnt{\small}

\newcommand{\xsat}{\mbox{\rm X3SAT}}

\title{On Salum's Algorithm for \xsat}

\author{Arian Nadjimzadah\thanks{Supported in part by NSF grant CCF-2006496.} \\
Department of Computer Science\\University of Rochester\\Rochester, NY 14627, USA
\and
David E. Narváez\thanks{Supported in part by NSF grant CCF-2030859 to the Computing Research Association for the CIFellows Project.}\\
Department of Computer Science\\University of Rochester\\ Rochester, NY
  14627, USA
 }

\date{April 6, 2021}

\newcommand{\cond}{\,\mid \:}

\newcommand{\sat}{\mbox{\rm SAT}}
\newcommand{\threesat}{\mbox{\rm 3SAT}}

\newcommand{\pe}{\mbox{\rm P}}

\newcommand{\np}{\mbox{\rm NP}}

\mathchardef\mhyphen="2D

\begin{document}
\sloppy
\maketitle

\begin{abstract}

This is a commentary on, and critique of, Latif Salum's paper titled 
``Tractability of One-in-three $\threesat$: $\pe = \np$.'' Salum purports to give a polynomial-time algorithm that solves the $\np$-complete problem $\xsat$, thereby claiming $\pe = \np$. The algorithm, in short, fixes the polarity of a variable, carries out simplifications over the resulting formula to decide whether to keep the value assigned or flip the polarity, and repeats with the remaining variables. One thing this algorithm does not do is backtrack. We give an illustrative counterexample showing why the lack of backtracking makes this algorithm flawed.

\end{abstract}

\section{Introduction}
We give a short commentary on Latif Salum's attempt at proving $\pe = \np$ in ``Tractability of One-in-three \threesat: $\pe = \np$''~\cite{salum2020tractability}, showing exactly where the argument breaks down. We first define the one-in-three $\threesat$ problem, also known as $\xsat$,
and its equivalence with Salum's nonstandard formulation.

The $\xsat$ problem takes a 3CNF formula (a conjunction of disjunctions, each having at most 3 literals) as input. A 3CNF formula $\phi$ is a \emph{yes} instance of $\xsat$ if and only if there exists an assignment that sets exactly one of the literals in every clause in $\phi$ to \emph{true}. $\xsat$ is known to be $\np$-complete.

Salum's paper introduces $\xsat$ \emph{formulas} (not to be confused with the $\xsat$ problem) of the form $\phi=\psi\land\varphi$, where $\psi$, called the \emph{minterm}, is a conjunction of literals, and $\varphi$ is a conjunction of $\xsat$ clauses of the form $(a \odot b \odot c)$.\footnote{Salum calls the $\odot$ symbol the ``exactly-1 disjunction'' which suggests it is an operator. Nevertheless, as an operator the semantics would be flawed as it is neither binary nor ternary. Instead, we use it in this critique as a purely notational element, which is essentially the role this symbol actually plays in Salum's paper.} An assignment of the variables in an $\xsat$ clause \emph{satisfies} the clause if and only if it sets exactly one of the literals $a$, $b$, or $c$ to $\emph{true}$. This holds similarly in the case of only two literals in $(a\odot b)$.
The \emph{minterm} $\psi$ is meant to hold a partial assignment of the literals in the formula and it is omitted when it is empty. An $\xsat$ formula $\phi$ is satisfiable if and only if there exists an assignment that makes $\psi$ \emph{true} and satisfies every clause in $\varphi$.

It is easy to see that $\xsat$ formulas with empty minterms are equivalent to 3CNF formulas (by simply replacing the $\odot$ notational element with $\lor$), 
and a satisfying assignment for an $\xsat$ formula with an empty minterm is also a witness for the equivalent 3CNF being a \emph{yes} instance of $\xsat$. Thus a polynomial-time algorithm to determine the satisfiability of $\xsat$ formulas would yield a polynomial-time algorithm for $\xsat$, thereby proving $\pe=\np$. Salum claims to provide such an algorithm.

Salum's algorithm~\cite{salum2020tractability} processes each variable in a given formula $\phi$ in turn, chooses a polarity, and simplifies the formula resulting from the chosen value. If a conflict is detected, the algorithm chooses the opposite polarity. We provide a more detailed summary in Section~\ref{sec:summary}. The fatal flaw in the algorithm is that decisions are not recorded to be later revised when a conflict is found. We provide a counterexample exploiting this flaw in Section~\ref{sec:counterexample}. We conclude in Section~\ref{sec:conclusions}.

\section{Summary of the Algorithm}
\label{sec:summary}

Salum's paper \cite{salum2020tractability} introduces much notation to define four functions that ultimately comprise the algorithm. These functions can be easily stated in plain English and through the use of standard Boolean formula satisfiability ($\sat$) terminology. Instead of reintroducing Salum's notation, we provide in Appendix~\ref{a:algorithms} line-by-line reinterpretations of the four key functions of the proposed algorithm, but using standard SAT terminology.

Recall, from standard SAT terminology and the description of the DPLL~\cite{dav-log-lov:j:machine-program-theorem-proving,dav-put:j:computing-proc-quant-theory} family of algorithms, that a \emph{decision}, during a solving process, corresponds to assigning an arbitrary truth value to a variable. The decision is then \emph{propagated} through the formula, i.e., the formula is simplified over the assignment made. During this process, the truth values of some of the other variables are set as consequences of the decision made. The literals thus set are typically called \emph{implied} literals. A solving process will typically simplify the formula iteratively until no further simplification can be achieved, at which point it will pick another variable to decide on and repeat. A decision should be recorded because the arbitrary nature of the assignment implies that it might be wrong, and the solving process would need to undo the decision and its consequences. Doing so is typically referred to as \emph{backtracking}. Backtracking is triggered when the current decisions have led to a situation where a variable must be set to both \emph{true} and \emph{false}, a situation typically referred to as a \emph{conflict}.

Salum's functions resemble many of the components of a typical DPLL algorithm, extended of course to the context of \xsat. $\mathtt{Reduce}(\phi_s,r_j)$ (Algorithm~\ref{alg:reduce}) propagates the decision $r_j$ over the formula $\phi_s$ and performs some simplifications. $\mathtt{Scope}(r_j,\phi_s)$ (Algorithm~\ref{alg:scope}) iteratively calls $\mathtt{Reduce}$ to propagate a single decision and all of its implied literals. Both $\mathtt{Reduce}$ and $\mathtt{Scope}$ return NULL when a conflict is detected. $\mathtt{Scan(\phi_s)}$ (Algorithm~\ref{alg:scan}), which is the entry point of Salum's algorithm, performs some simplifications, then decides on a variable that is still unassigned in the simplified formula, and calls $\mathtt{Scope}$ on this decision. If $\mathtt{Scope}$ detects a conflict due to this decision, $\mathtt{Scan}$ calls $\mathtt{Remove}$ (Algorithm~\ref{alg:remove}) to assign the literal to the opposite polarity. $\mathtt{Remove}(r_j,\phi_s)$ calls $\mathtt{Reduce}(\phi_s,\overline{r_j})$ but this time, if a conflict is detected, it returns UNSAT\@. If no conflict is detected, $\mathtt{Remove}(r_j, \phi_s)$ simplifies the current formula over the assignment $\overline{r_j}$ and calls $\mathtt{Scan}(\phi_{s+1})$ over the resulting formula $\phi_{s+1}$.

Our analysis reveals that the proposed algorithm is missing one key component of DPLL algorithms: it does not backtrack. Without backtracking, it is usually easy to show that a solving algorithm is incorrect by providing a formula and an ordering of the literals that trigger a bad decision. We do that in Section~\ref{sec:counterexample}.
\section{Illustrative Counterexample}
\label{sec:counterexample}

We provide a counterexample that ``tricks'' \texttt{Scope}, namely,
\begin{equation*}
    \phi = (a \odot b \odot c) \land (b \odot x \odot y) \land (c \odot x \odot \bar y).
\end{equation*}

This formula is quite clearly satisfiable, for instance, by the assignment 
$(a,b,c,x,y) = (0,0,1,0,1)$.\footnote{As is standard, 1 represents \emph{true} and 0 represents \emph{false}.}
Salum claims in~\cite[Section 3.1]{salum2020tractability} that if $\mathtt{Scan}(\phi)$ does not find a satisfying assignment (in Salum's words, it is ``interrupted''), then $\phi$ is unsatisfiable, so we run $\mathtt{Scan}(\phi)$ and show it does not find a satisfying assignment. The line numbers mentioned in the explanation below correspond to the line numbers of the algorithms in the original paper~\cite{salum2020tractability}, and not to those of our reinterpretations in Appendix~\ref{a:algorithms}.

The \emph{minterm} in $\phi$ is empty, so we skip to line 4 of $\mathtt{Scan}$ to process the literals in turn. Salum claims in \cite[Theorem 36]{salum2020tractability} that the order in which the literals are processed is arbitrary. Accordingly, we choose to process $a$ first, invoking $\mathtt{Scope}(a, \phi)$ in line 6. From line 4 of $\mathtt{Scope}(a, \phi)$, we call $\mathtt{Reduce}(\phi, a)$. Given that $a$ was chosen to be \emph{true}, the first clause in $\phi$ propagates to the minterm $a \land \bar b \land \bar c$.
The second and third clauses contain neither $a$ nor $\bar a$ so they are not touched. Thus at the end of $\mathtt{Reduce}(\phi, a)$, we are left with 
\begin{equation*}
    \phi = (a \land \bar b \land \bar c) \land (b \odot x \odot y) \land (c \odot x \odot \bar y).
\end{equation*}

Back in $\mathtt{Scope}(a,\phi)$, line 2 will call $\mathtt{Reduce}(\phi, \bar b)$ and $\mathtt{Reduce}(\phi, \bar c)$ in turn. $\mathtt{Reduce}(\phi, \bar b)$ will first propagate the decision $\bar b$. Since there are no clauses containing $\bar b$ it moves on to line 8, where it will delete $\overline{\overline b} = b$ from all clauses containing it. The only such clause is $(b \odot x \odot y)$, and by removing $b$ we obtain $(x \odot y)$. When we return to line 3 of $\mathtt{Scope}(a,\phi)$, the current formula is
\begin{equation*}
    \phi = (a \land \bar b \land \bar c) \land (x \odot y) \land (c \odot x \odot \bar y).
\end{equation*}

Analogously, in $\mathtt{Reduce}(\phi, \bar c)$ we reduce $(c \odot x \odot \bar y)$ to $(x \odot \bar y)$, thus obtaining the formula
\begin{equation*} 
    \phi = (a \land \bar b \land \bar c) \land (x \odot y) \land (x \odot \bar y),
\end{equation*}
upon finishing $\mathtt{Scope}(a, \phi)$.

At this point the algorithm is already doomed as, having chosen the positive polarity of $a$, it has no way to satisfy the remainder of the formula, namely $(x \odot y) \land (x \odot \bar y)$. For completeness, however, we continue with the remainder of the execution. 

Having returned to line 6 of $\mathtt{Scan}(\phi)$, we now run $\mathtt{Scope}(\bar a,\phi)$.  Line 5 in $\mathtt{Scope}(\bar a,\phi)$ detects the conflict with $a$ already in the minterm, and returns NULL\@.
Thus $\mathtt{Scan}(\phi)$ calls $\mathtt{Remove}(\bar a, \phi)$, which leaves the formula unaffected, as neither $a$ nor $\bar a$ is contained in the remainder of the formula $(x \odot y) \land (x \odot \bar y)$.

The execution returns to line 4 of $\mathtt{Scan}(\phi)$, and we process the variable $x$ next. Now in $\mathtt{Scope}(x, \phi)$, we simplify the formula $\phi$ due to $x$ being assigned to \emph{true}, and obtain 
\begin{equation*}
    \phi = (a \land \bar b \land \bar c \land x \land \bar y \land y),
\end{equation*}
which is detected in line 6 of $\mathtt{Reduce}(\phi,x)$ as a conflict. Thus $\mathtt{Scan}(\phi)$ calls $\mathtt{Remove}(\phi,x)$ which in turn calls $\mathtt{Reduce}(\phi, \bar x)$. $\mathtt{Reduce}$ then detects the conflict $y \land \bar y$ and returns NULL\@. $\mathtt{Remove}(\phi,x)$ thus concludes that the formula is unsatisfiable. In the next subsection, we provide reasoning for why this counterexample ``tricks'' Salum's algorithm.

\subsection{Key Issue with the Proposed Algorithm}

As mentioned at the end of Section~\ref{sec:summary}, Salum's algorithm \cite{salum2020tractability} fails to backtrack.
Our example exploits this by providing a formula in which, after setting the value of variable $a$ to true, simplification results in an unsatisfiable formula.

It is certainly the case that picking a different variable might lead to a satisfying assignment using the proposed algorithm. However, requiring a particular ordering of the variables contradicts the claim of monotonicity~\cite[Theorem 36]{salum2020tractability} and several assertions in the paper stating that the order of the decisions is arbitrary.
One could think of several simple ways to fix this, leading to heuristics about how to pick the next literal. We discuss a few, for which it is easy to extend our counterexample:

\begin{itemize}
    \item Picking a variable based on how many clauses contain the variable: In this case, we add enough clauses of the form $(a\odot x_i\odot y_i)$ to make $a$ the most frequent variable.
    \item Reversing the order of the polarities in Line~5 of $\mathtt{Scan}$: In the example in Section \ref{sec:counterexample}, we chose to first process the positive literal of $a$. The proposed algorithm does not specify the order in which to process $r_i\in\{x_i,\overline{x_i}\}$.
    One could argue that our counterexample would not work had we picked the negative literal first, and that perhaps a fixed order of negative literal, then positive literal, would resolve the issue. In this case, we construct the same formula but with $\overline{a}$ in the place of $a$.
    \item Picking a variable in reverse lexicographical order: We labeled our variables such that the first (in lexicographical order) variable leads to a bad decision. Suppose one wanted to fix this issue by reversing the order of the variables. Then a simple relabeling still yields a counterexample.
\end{itemize}

This discussion suggests that the fundamental issue of the proposed algorithm is likely not addressable by simple fixes. In essence, simply propagating each decision is not enough to preclude the need for backtracking. That is, the algorithm can only ever backtrack a single step by flipping the polarity of the literal just chosen. The only obvious way to remedy this is to allow for proper backtracking, which in principle will make the algorithm correct, but potentially at the enormous cost of exponential worst-case running time.

\section{Conclusions}
\label{sec:conclusions}

We have studied Latif Salum's paper titled ``Tractability of One-in-three $\threesat$: $\pe = \np$''~\cite{salum2020tractability}. We showed that this algorithm is flawed by providing a counterexample: an \xsat{} formula that is satisfiable, but for which the algorithm returns UNSAT\@. Thus Salum's proposed algorithm does not settle the long-standing question of $\pe$ vs.\ $\np$. In fact, an algorithm as straightforward as the one proposed will very likely fail to show $\pe=\np$.

\section*{Acknowledgments}
We thank Michael C. Chavrimootoo, Lane A. Hemaspaandra, and Mandar Juvekar for their helpful feedback on an earlier version of our critique. Any remaining errors are the responsibility of the authors.

\bibliography{citations}

\begin{thebibliography}{DLL62}

\bibitem[DLL62]{dav-log-lov:j:machine-program-theorem-proving}
M.~Davis, G.~Logemann, and D.~Loveland.
\newblock A machine program for theorem-proving.
\newblock {\em Communications of the ACM}, pages 394--397, July 1962.

\bibitem[DP60]{dav-put:j:computing-proc-quant-theory}
M.~Davis and H.~Putnam.
\newblock A computing procedure for quantification theory.
\newblock {\em Journal of the ACM}, 7(3):201--215, 1960.

\bibitem[Sal20]{salum2020tractability}
Latif Salum.
\newblock Tractability of one-in-three {3SAT}: {P = NP}.
\newblock Technical Report arXiv:2012.06304v1~[cs.CC], Computing Research
  Repository, \mbox{arXiv.org/corr/}, December 2020.
\newblock Version 1.

\end{thebibliography}

\newpage 

\begin{appendix}
\section{Algorithms}
\label{a:algorithms}

The algorithms in this appendix do not replace those in the original paper~\cite{salum2020tractability}, but are instead meant to help the reader quickly grasp the nature of the algorithms therein. In particular, in Section~\ref{sec:counterexample}, when we refer to line numbers, we refer to those of the algorithms in the original paper. Our reinterpretations use standard SAT terminology, but we mix in some of Salum's notation when appropriate in order to provide an easy way for the reader to map the original expositions of the algorithms to our reinterpretations.

\begin{algorithm}[b!]
\DontPrintSemicolon
\caption{$\mathtt{Reduce}(\phi_s,r_j)$}
\label{alg:reduce}
\Input{$\xsat$ formula $\phi_s$ and literal $r_j$}
\Output{Partial assignment $\tilde{\psi_s}$ and a simplified formula $\phi'^{\overline{r_j}}$}
\Begin{
\For{every clause $C_k$ that contains $r_j$}{
Add the propagation of $r_j$ in $C_k$ to $\tilde{\psi_s}$, i.e., add the literals $\{r_j\}\cup\{\overline{r_i}\cond r_i\in C_k \land r_i\neq r_j\}$\;
\If{a conflict is detected}{
\Return{\texttt{NULL}}
}
}
\For{every clause $C_k$ where $\overline{r_j}$ appears}{
Delete $\overline{r_j}$ from $C_k$\;
\If{$C_k$ turns into a unit clause}{
Add the unique literal in $C_k$ to $\tilde{\psi_s}$, remove $C_k$ from the formula $\phi'^{\overline{r_j}}$\;
}
\If{conflict is detected}{
\Return{\texttt{NULL}}
}
}
}
\end{algorithm}

\begin{algorithm}[b!]
\DontPrintSemicolon
\caption{$\mathtt{Scope}(r_j,\phi_s)$}
\label{alg:scope}
\Input{Literal $r_j$ and $\xsat$ formula $\phi_s$}
\Output{Partial assignment $\psi(r_j)$ and simplified formula $\phi'_s(r_j)$}
\Begin{
Initialize $\psi(r_j)\gets \{r_j\}$, $\phi'_s(r_j)\gets \phi_s$\;
\For{all unprocessed literals $r_e$ in the current partial assignment}{
\If{a conflict is detected in $\mathtt{Reduce}(\phi'_s(r_j),r_e)$}{
\Return{\texttt{NULL}}
}
Add the partial assignment from the $\mathtt{Reduce}$ call to the current assignment $\psi(r_j)$\;
\If{a conflict is detected}{
\Return{\texttt{NULL}}
}
Replace the current simplified formula $\phi'_s(r_j)$ with the one obtained from $\mathtt{Reduce}$\;
Mark $r_e$ as processed\;
}
}
\end{algorithm}

\begin{algorithm}[b!]
\DontPrintSemicolon
\caption{$\mathtt{Scan}(\varphi_s)$}
\label{alg:scan}
\Input{An $\xsat$ formula $\varphi_s$}
\Output{Partial assignment $\hat{\psi}$ and simplified function $\hat{\phi}$}
\Begin{
\For{every literal $r_i$ still in $\phi$ such that $\overline{r_i}$ is in the current partial assignment}{
$\mathtt{Remove}(r_i,\phi_s)$\;
}
\For{every literal $x_i$ still in the formula}{
\For{every polarity $r_i\in\{x_i,\overline{x_i}\}$}{
\If{$\mathtt{Scope}(r_i,\phi_s)$ detects a conflict}{
$\mathtt{Remove}(r_i,\phi_s)$\;
}
}
}
}
\end{algorithm}

\begin{algorithm}[b!]
\DontPrintSemicolon
\caption{$\mathtt{Remove}(r_j,\phi_s)$}
\label{alg:remove}
\Input{Literal $r_j$ and $\xsat$ formula $\phi_s$}
\Begin{
\If{$\mathtt{Reduce}(\phi_s,\overline{r_j})$ detects a conflict}{
\Return{UNSAT}
}
Update the current partial assignment with the information from $\mathtt{Reduce}$\;
\If{a conflict is detected}{
\Return{UNSAT}
}
Remove $r_j$ from the list of literals in the formula, add it to the list of literals in the partial assignment\;
The new formula $\phi_{s+1}$ is calculated by simplifying $\phi_s$ over $\overline{r_j}$\;
$\mathtt{Scan}(\phi_{s+1})$\;
}
\end{algorithm}

\end{appendix}

\end{document}